\documentclass[tradiabstract]{aa} 
\usepackage{graphicx}
\usepackage{supertabular}
\usepackage{txfonts}
\begin{document}
\title{Discovery of close binary central stars in the planetary nebulae NGC~6326 and NGC~6778\thanks{Based on observations made with Gemini South under program GS-2009A-Q-35, the South African Astronomical Observatory 1.9 m telescope, the Flemish Mercator telescope of the Observatoria del Roque de Los Muchachos, and the Very Large Telescope at Paranal Observatory under program 085.D-0629(A).}} 
   \author{B. Miszalski
           \inst{1,2,3}
           \and
           D. Jones
           \inst{4}
           \and
           P. Rodr\'iguez-Gil
           \inst{5,6}
           \and
           H. M. J. Boffin
           \inst{4}
           \and
           R. L. M. Corradi
           \inst{5,6}
           \and
           M. Santander-Garc\'ia
           \inst{7,8,5}
          }
\institute{Centre for Astrophysics Research, STRI, University of Hertfordshire, College Lane Campus, Hatfield AL10 9AB, UK
\and
South African Astronomical Observatory, PO Box 9, Observatory, 7935, South Africa
\and
Southern African Large Telescope Foundation, PO Box 9, Observatory, 7935, South Africa\\
\email{brent@saao.ac.za}
\and
European Southern Observatory, Alonso de Cordova 3107, Casilla 19001, Santiago, Chile
\and
Instituto de Astrof\'isica de Canarias, E-38200 La Laguna, Tenrife, Spain
\and
Departmento de Astrof\'isica, Universidad de La Laguna, E-38205 La Laguna, Tenerife, Spain
\and
Observatorio Astron\'omico Nacional, Ap 112, 28803 Alcal\'a de Henares, Spain
\and
CAB, INTA-CSIC, Ctra de Torrej\'on a Ajalvir, km 4, 28850 Torrej\'on de Ardoz, Madrid, Spain
         }
   \date{Received -; accepted -}

   \abstract{
   We present observations proving the close binary nature of the central stars belonging to the planetary nebulae (PNe) NGC~6326 and NGC~6778. Photometric monitoring reveals irradiated lightcurves with orbital periods of 0.372 and 0.1534 days, respectively, constituting firm evidence that they passed through a common-envelope (CE) phase. Unlike most surveys for close binary central stars (CSPN) however, the binary nature of NGC~6326 was first revealed spectroscopically and only later did photometry obtain an orbital period. Gemini South observations revealed a large 160 km/s shift between the nebula and emission lines of C~III and N~III well known to originate from irradiated atmospheres of main-sequence companions. These so-called weak emission lines are fairly common in PNe and measurement of their radial velocity shifts in spectroscopic surveys could facilitate the construction of a statistically significant sample of post-CE nebulae. There is growing evidence that this process can be further accelerated by pre-selecting nebulae with traits of known post-CE nebulae. Both NGC~6326 and NGC~6778 were selected for their rich attribution of low-ionisation filaments and collimated outflows, thereby strengthening the connection between these traits and post-CE CSPN. 
   }
   \maketitle
   \section{Introduction}
   About 40 binary central stars of planetary nebulae (CSPN) are now known (Bond 2000; De Marco, Hillwig \& Smith 2008; Miszalski et al. 2009a, 2011a). The overwhelming majority of which were discovered by measuring periodic photometric variations due to irradiation, ellipsoidal modulation and eclipses. Their short orbital periods of $P_\mathrm{orb}$$\sim$0.1--1.0 days (Miszalski et al. 2009a, 2011a) are firm evidence that they passed through a common-envelope (CE) phase (Iben \& Livio 1993). During the CE phase a strong equatorial density contrast is expected to develop and sculpt highly asymmetric nebulae (Sandquist et al. 1998). The degree to which this may occur is highly uncertain and other mechanisms may be responsible for shaping the extraordinary diversity of PNe morphologies (Balick \& Frank 2002). A binary companion is emerging as the preferred method for shaping PNe (Soker 2006; Nordhaus, Blackman \& Frank 2007), but their full effect can only be elucidated if a statistically significant sample of post-CE PNe is accumulated and studied in detail. 
   
   Careful sample selection is essential to maximise the chances of finding new post-CE CSPN as some nebula features have been tied exclusively to binaries. Soker \& Livio (1994) concluded that collimated polar outflows or jets could only be powered by a post-CE CSPN and there are many examples where multiple pairs of jets may be the result of precessing jets (see e.g. Sahai 2000; Sahai, Morris \& Villar 2011). The first steps towards proving the connection between jets and post-CE CSPN have been made (Mitchell et al. 2007; Corradi et al. 2011; Miszalski et al. 2011b), and further post-CE specific trends have been identified (Miszalski et al. 2009b), but many other challenges remain (De Marco 2009). Further progress depends critically on developing new survey techniques for binary CSPN to build a statistically significant sample of a few hundred post-CE nebulae. Single object photometric monitoring has been the most commonly used strategy (e.g. Bond 2000; Hajduk et al. 2010; Hillwig et al. 2010), although it is not the best way to reach our goals given the large time expenditure involved with following orbital periods $P_\mathrm{orb}$ of $\sim$0.1--1.0 days. Modern microlensing surveys essentially remove this barrier (Miszalski et al. 2009a; Lutz et al. 2010) and can rapidly build up samples of post-CE nebulae for study (Miszalski et al. 2009b). Alternative strategies are still required to discover close binaries in PNe outside microlensing surveys and those with brighter nebulae not amenable to photometric monitoring (especially in poor seeing).
  
   In this paper we demonstrate the potential for large-scale spectroscopic surveys to discover close binaries in less time than traditional photometric surveys. Section \ref{sec:obs} presents proof of close binary CSPN in NGC~6326 (PN G338.1$-$08.3) and NGC~6778 (PN G034.5$-$06.7), both selected for study solely because of their rich attribution of low-ionisation structures (Schwarz, Corradi \& Melnick 1992; Corradi et al. 1996; Miranda et al. 2010). These discoveries were made during the course of ongoing work by our team (Miszalski et al. 2009a, 2011a, 2011b; Corradi et al. 2011; Santander-Garc\'ia et al. 2011) that aims to further strengthen the trends specific to post-CE nebulae first identified by Miszalski et al. (2009b). Section \ref{sec:discussion} discusses the morphological features of both nebulae and Sect. \ref{sec:conclusion} concludes.  

   \section{Observations}
   \label{sec:obs}
   \subsection{Gemini South GMOS spectroscopy of NGC~6326}
\begin{figure}
      \begin{center}
         \includegraphics[scale=0.4,angle=270]{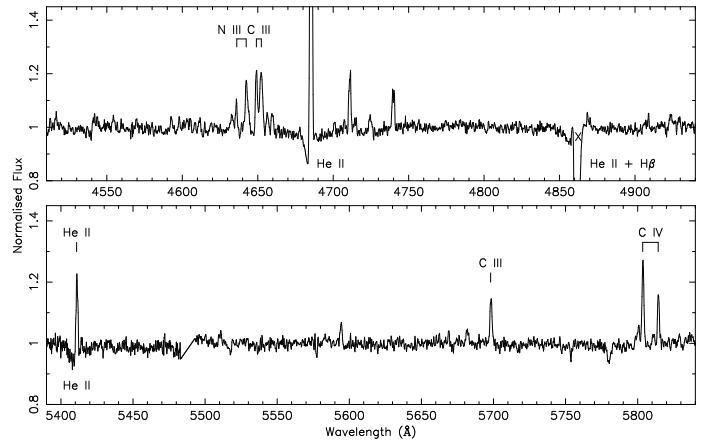}
      \end{center}
      \caption{Gemini South GMOS rectified spectrum of NGC~6326.}
      \label{fig:gmosN6326}
   \end{figure}

   Central star magnitudes of NGC~6326 were estimated by Shaw \& Kaler (1989) to be $B=16.08\pm0.35$ and $V=15.53\pm0.27$ mag, and Tylenda et al. (1991) to be $B=16.3\pm0.5$ and $V=16.1\pm0.5$ mag. During the Gemini South program GS-2009A-Q-35 on 4 April 2009 we obtained $2\times1800$ s longslit spectra of the CSPN with the B1200 grating and 0.75\arcsec\ slit of GMOS (Hook et al. 2004). A total wavelength range of 4085--5977 \AA\ was observed at a resolution of 1.6 \AA\ (full width at half-maximum, FWHM) and a dispersion of 0.23 \AA/pixel. The data were reduced with the Gemini \textsc{iraf} package and adjacent nebula emission was subtracted from extracted CSPN spectra at a spatial resolution of 0.146\arcsec/pixel.  Figure \ref{fig:gmosN6326} shows the second of two spectra whose stellar emission lines of C~III, C~IV and N~III were found to be redshifted $\sim$160 km/s and $\sim$210 km/s compared to the nebula and He~II absorption lines from the primary, respectively. 

   Ordinarily the C and N emission lines would be sufficient for a so-called weak emission line or \emph{wels} CSPN classification (Tylenda, Acker \& Stenholm 1993), but the large radial velocity (RV) shifts strongly suggested a spectroscopic post-CE binary which we confirm in Sects. \ref{sec:lc} and \ref{sec:fors}. Miszalski et al. (2011b) were the first to suggest that many CSPN classified as \emph{wels} should turn out to be post-CE binaries with more specialised observations (see also Sect. \ref{sec:conclusion}). Pollacco \& Bell (1993, 1994) showed the C and N lines trace the irradiated zone of the main-sequence companion and this explains their high level of variability (e.g. Sect. \ref{sec:fors}). Wawrzyn et al. (2009) suggested the enriched C and N abundance required to reproduce the strong lines may be the result of polluted material accreted from the primary. NGC~6326 is also one of a few post-CE binaries showing the higher ionisation C~IV lines (Pollacco \& Bell 1994; Corradi et al. 2011; Miszalski et al. 2011b). 

   \subsection{Lightcurves of NGC~6326 and NGC~6778}
   \label{sec:lc}
\begin{figure}
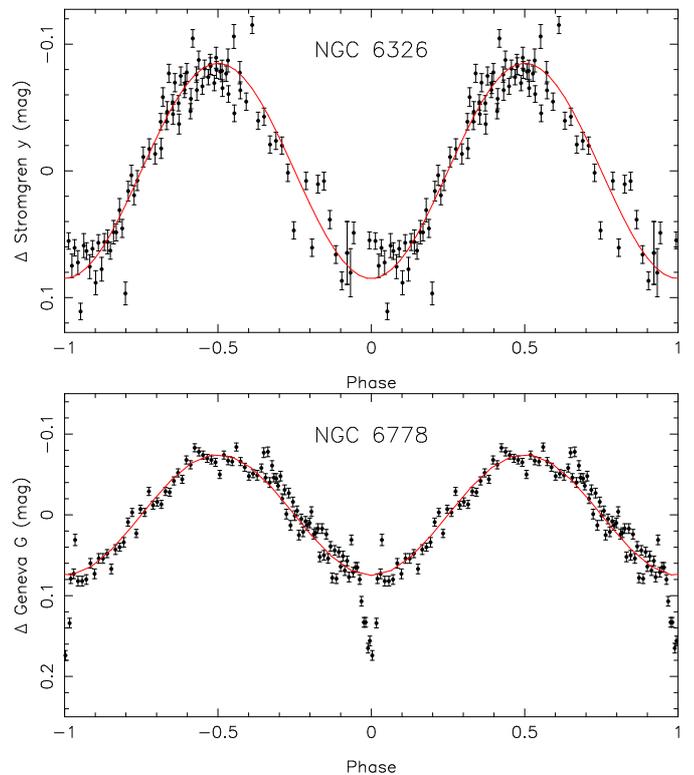

      \begin{center}
         \includegraphics[scale=0.50]{NGC6326-phased.ps}\\
         \includegraphics[scale=0.50]{NGC6778-phased.ps}
      \end{center}
      \caption{Lightcurves of NGC~6326 and NGC~6778 with sinusoidal fits of amplitudes 0.085 mag (NGC~6326) and 0.075 mag (NGC~6778).}
      \label{fig:lc}
   \end{figure}

   \emph{NGC~6326} was observed with the SAAO 1.9 m on the nights 11, 13, 16 and 17 August 2010. The SAAO CCD camera was used to obtain 10 min exposures through a Str\"omgren $y$ filter ($\lambda_\mathrm{eff}=547$ nm, $W_0=23$ nm) with a 1K$\times$1K STE4 CCD in direct imaging mode. The pixel scale was 0.14\arcsec/pix and the field-of-view $146\times146$ arcsec$^2$. Four continuous blocks totalling 14.72 hours were observed across the nights for 2.73, 1.90, 5.20 and 4.89 hours, respectively. While the strongest nebula lines were removed by the Str\"omgren $y$ filter, the nebula continuum remained to produce an image similar to the GMOS OIIIC image (Sect. \ref{sec:image}; see also Ciardullo \& Bond 1996). To minimise the influence of the remaining nebula we used a fixed aperture of 3.5\arcsec\ radius that is $\sim$1.5 times larger than the poorest seeing during the observations (Naylor 1998; Jones 2011). Figure \ref{fig:lc} shows the lightcurve phased with the ephemeris HJD (min $y$) = 2455425.713 $+$ 0.372$\pm0.002$ $E$. No correlation was found between the photometry extracted with \textsc{sextractor} (Bertin \& Arnouts 1996) and the seeing, thereby ruling out nebula contamination as the source of the variability (Jones 2011). There is however a residual level of scatter probably introduced by the bright nebula continuum, but the effect is much less than the sinusoidal variability due to orbital motion. 

   \emph{NGC~6778} was observed for a continuous 6.35 hours on 19 June 2010 with the MEROPE camera (Davignon et al. 2004) on the Flemish 1.2 m Mercator telescope (Raskin et al. 2004). A Geneva $G$ filter ($\lambda_\mathrm{eff}=576.59$ nm, $W_0=22.18$ nm) was used to help minimise the nebula contamination. The long continuous monitoring using 3 min exposures allowed the same eclipse to be observed twice and ellipsoidal modulation to be ruled out. Figure \ref{fig:lc} shows the lightcurve phased with the ephemeris HJD (min $G$) = 2455365.715 $+$ 0.1534$\pm$0.0001 $E$. The short orbital period of 3.68 hours was previously reported by us in Miszalski et al. (2011a) and is one of the shortest known in PNe. 
   
   \subsection{Additional spectroscopy of NGC~6326 and NGC~6778}
   \label{sec:fors}
\begin{figure}
      \begin{center}
         \includegraphics[scale=0.4,angle=270]{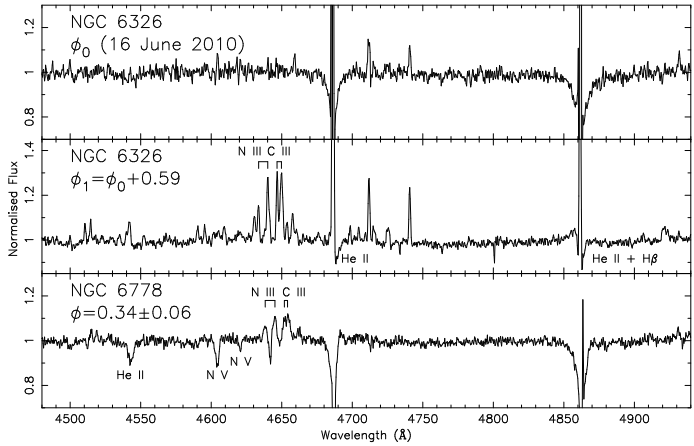}
      \end{center}
      \caption{VLT FORS2 rectified spectra of NGC~6326 and NGC~6778.}
      \label{fig:fors}
   \end{figure}

   We also observed both CSPN with FORS2 (Appenzeller et al. 1998) on the VLT under program ID 085.D-0629(A). The instrumental setup is the same as that used by Miszalski et al. (2011b). On 16 June 2010 a 1500 s exposure of NGC~6778 was taken with a 0.7\arcsec\ slit, followed by an 1800s exposure of NGC~6326 with a 0.5\arcsec\ slit. An identical exposure of NGC~6326 was taken the next night ($\Delta\phi=\phi_1-\phi_0=0.59$). Figure \ref{fig:fors} displays the reduced and rectified spectra. Unfortunately the large uncertainty in the period of NGC~6326 means accurate individual phases for the FORS2 spectra taken $\sim$2 months earlier cannot be calculated. It is clear however that the first spectrum was taken near minimum light where the irradiated atmosphere of the secondary is hidden, while the second spectrum was taken near light maximum where the irradiated C~III and N~III emission now dominates the spectrum. This striking change between nights is not uncommon amongst close binary CSPN (e.g. Exter et al. 2005; Wawrzyn et al. 2009) and we discuss the influence of this effect on surveys for binary CSPN in Sect. \ref{sec:conclusion}. The phase for NGC~6778 was calculated to be $0.34\pm0.06$, where the uncertainty reflects the exposure time. Note this value is reliable since the FORS2 spectrum was taken within a few days of the lightcurve observations. Again we see the presence of C~III and N~III emission consistent with the irradiated lightcurve in Fig. \ref{fig:lc}.
   
   \subsection{Imaging of NGC~6326 and NGC~6778}
   \label{sec:image}
\begin{figure}
      \begin{center}
         \includegraphics[scale=0.3]{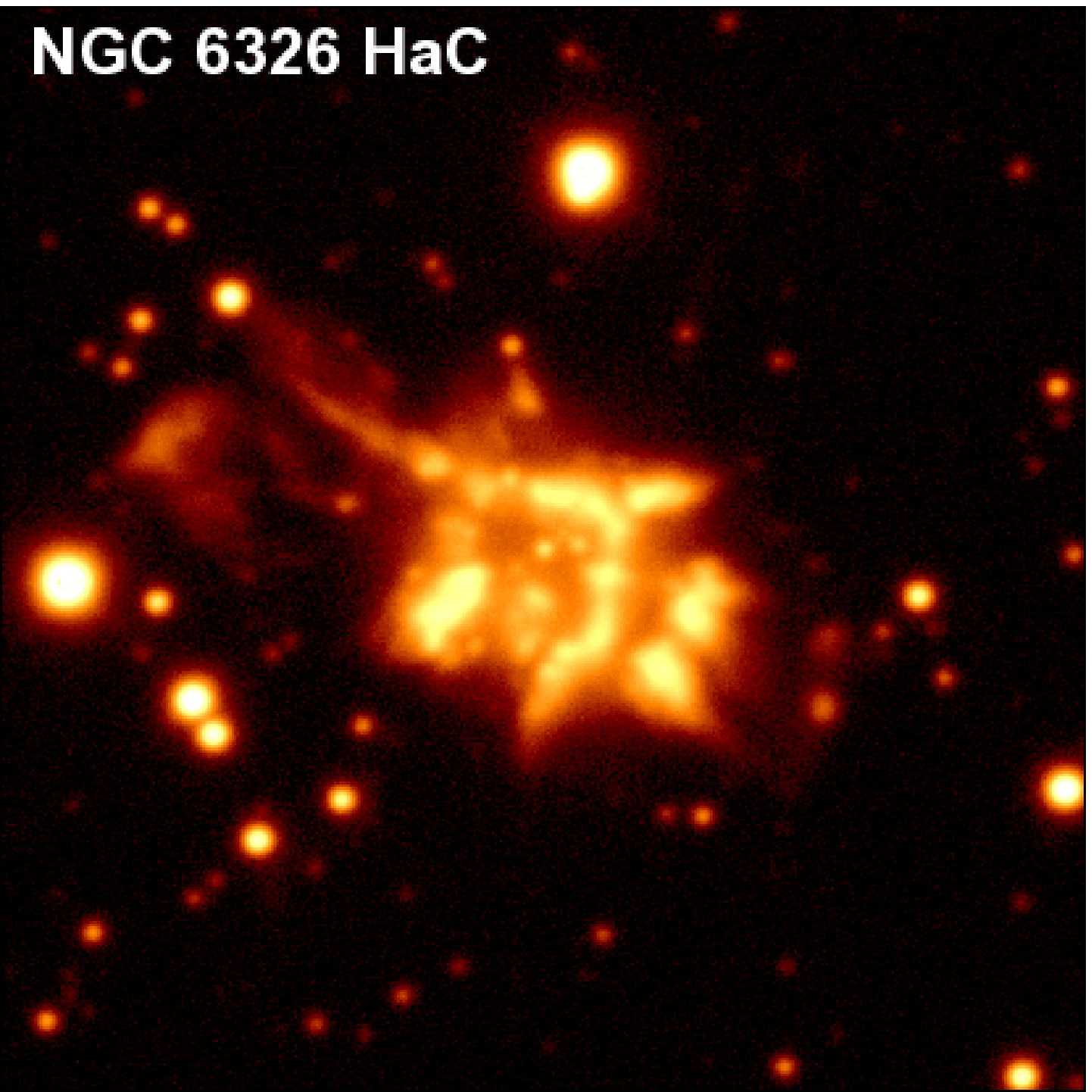}
         \includegraphics[scale=0.3]{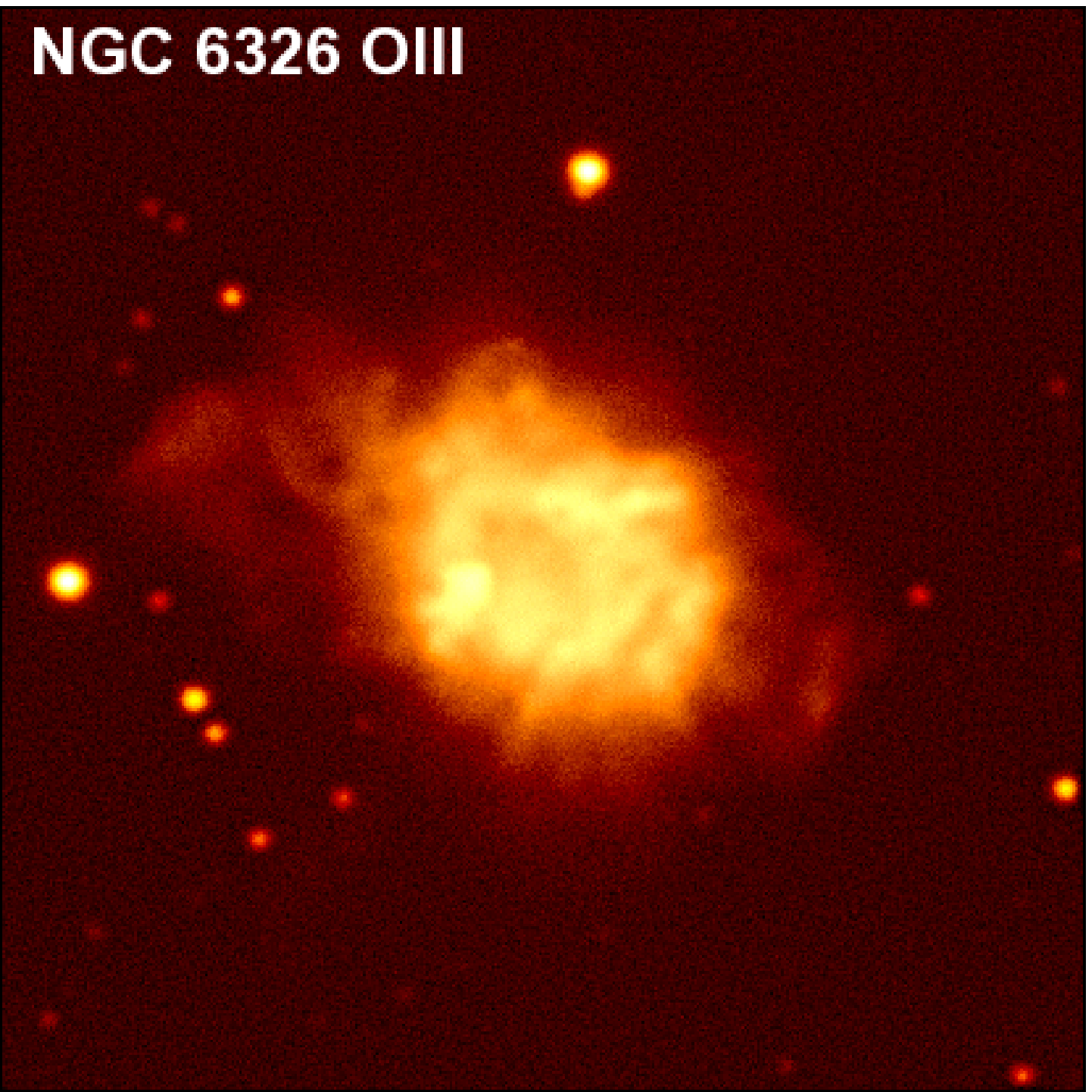}\\
         \includegraphics[scale=0.3]{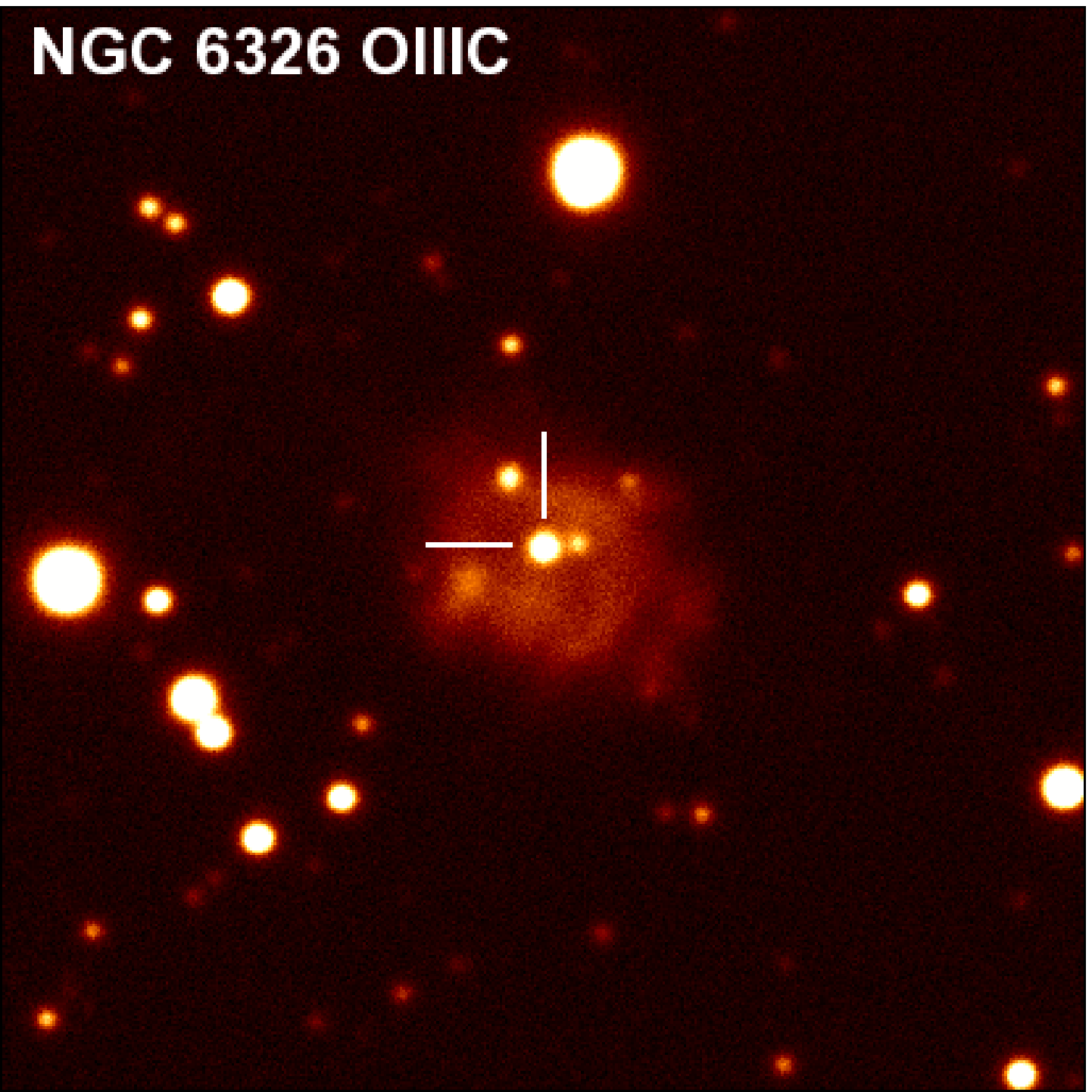}
         \includegraphics[scale=0.3]{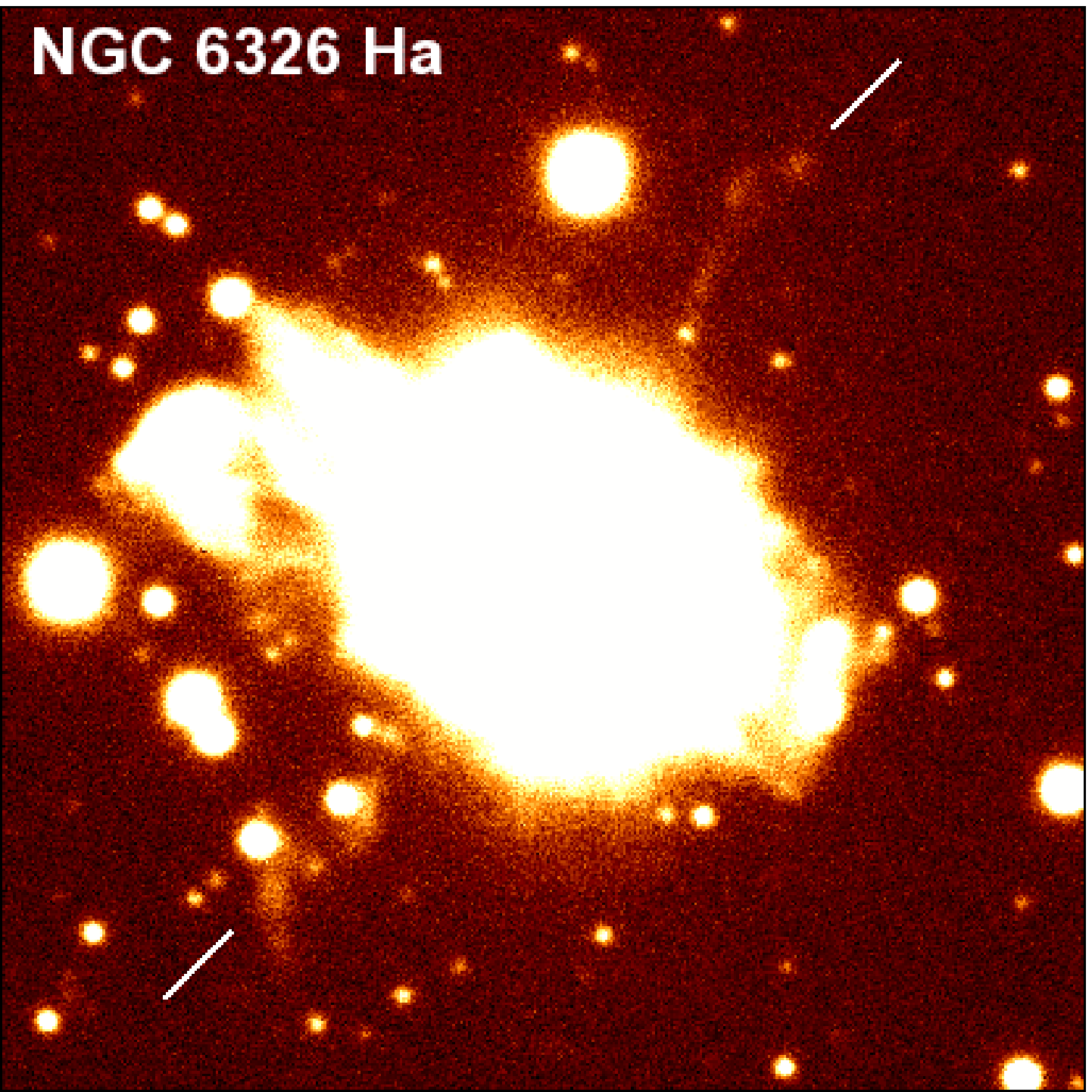}\\
         \includegraphics[scale=0.38895]{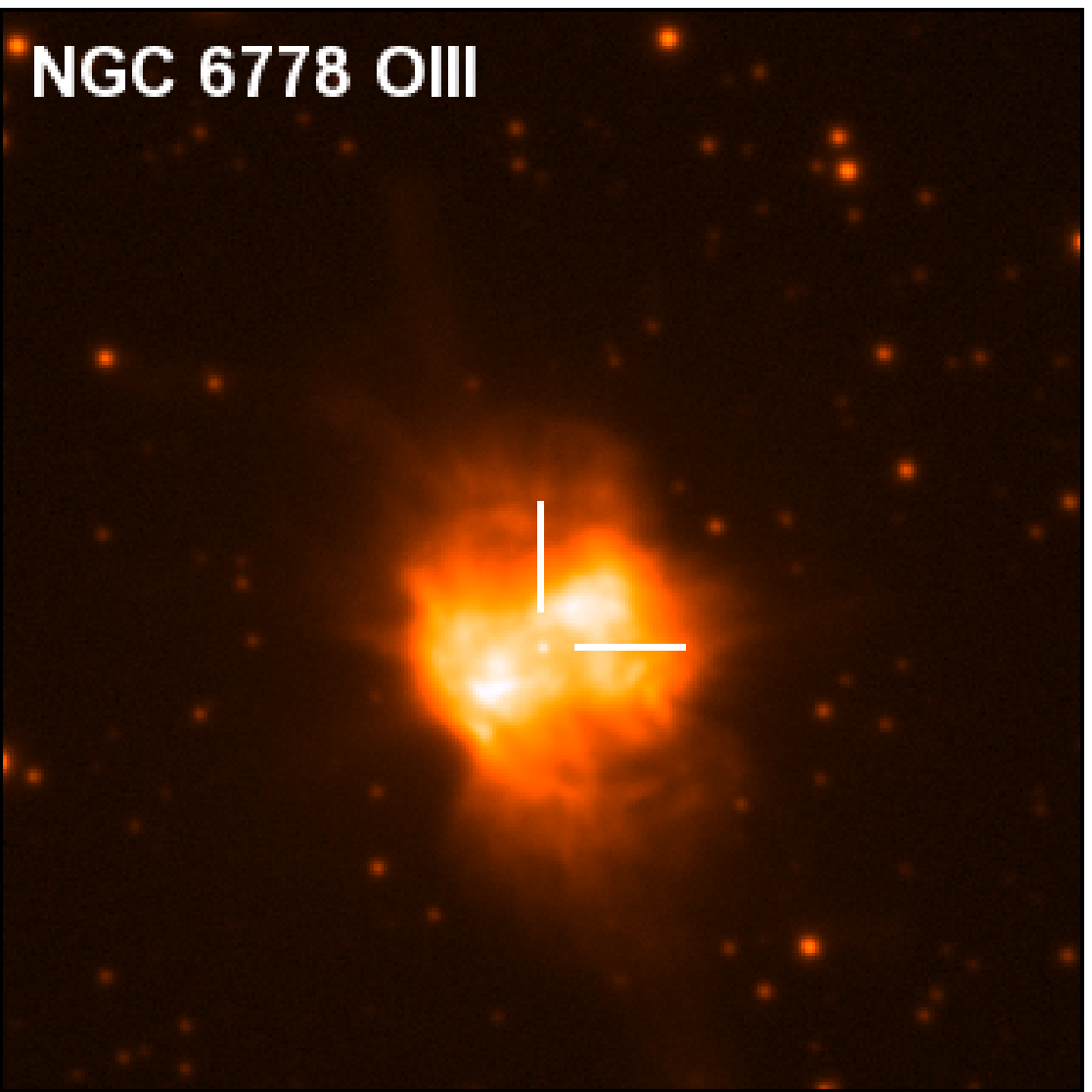}
         \includegraphics[scale=0.38895]{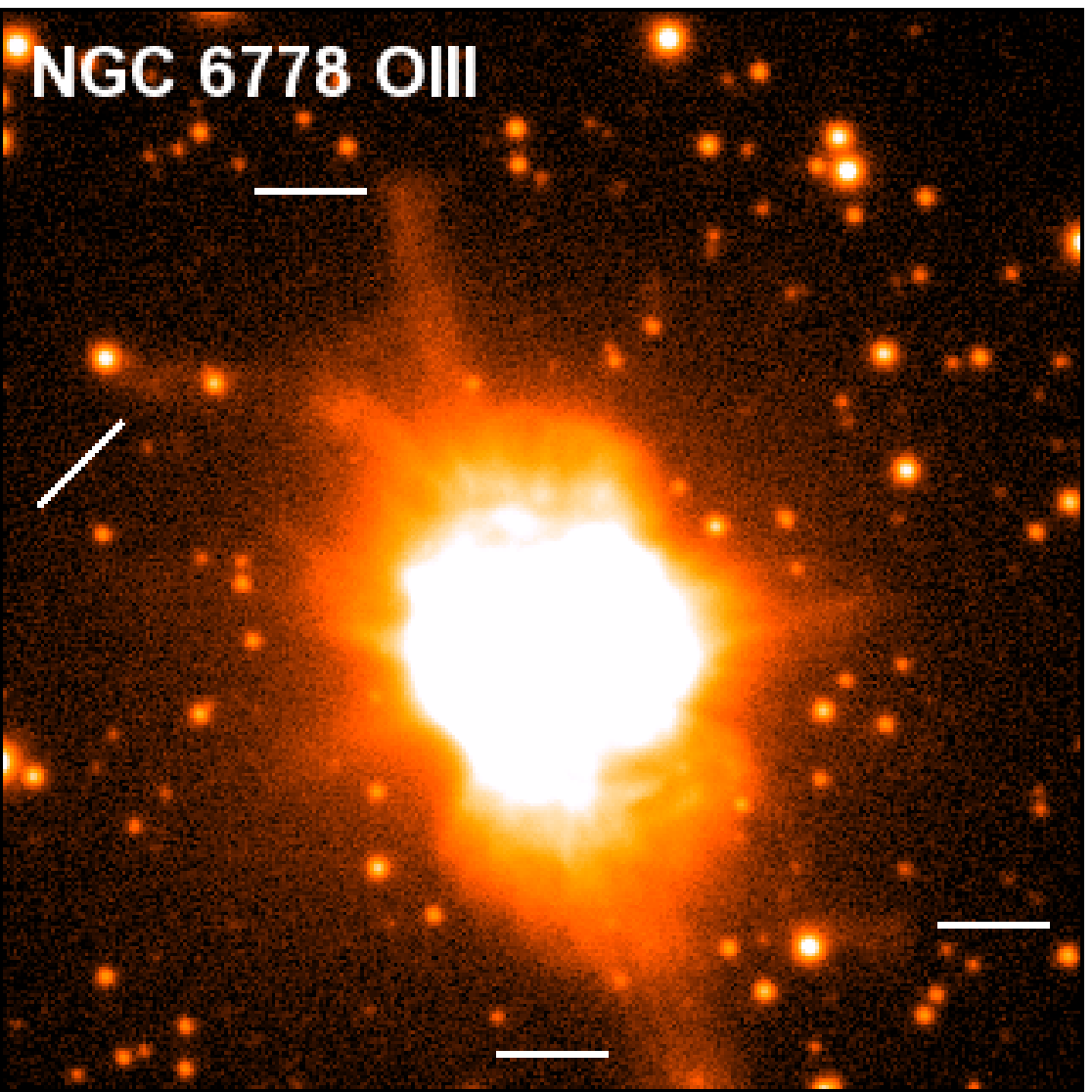}
      \end{center}
      \caption{GMOS images of NGC~6326 in [N~II] (HaC), [O~III] (OIII), [O~III]-continuum (OIIIC) and H$\alpha$+[N~II] (Ha), and VLT FORS2 images of NGC~6778 in [O~III]. Lines point to the CSPN positions and the tips of collimated outflows. Images are $60\times60$ arcsec$^2$ (NGC~6326) and $80\times80$ arcsec$^2$ (NGC~6778) with North up and East to the left. 
      }
      \label{fig:montage}
   \end{figure}

   Figure \ref{fig:montage} is a montage of images observed by us of NGC~6326 and NGC~6778. NGC~6326 was imaged during GMOS acquisition with exposures taken in the HaC filter (120 s) and Ha, OIII and OIIIC filters (60 s). The central wavelength and FWHM of the filters are 499.0/4.5 nm (OIII), 514.0/8.8 nm (OIIIC), 656.0/7.2 nm (Ha), 662.0/7.1 nm (HaC) and 672.0/4.4 nm (SII). The Ha filter includes H$\alpha$ and both [N~II] lines, while the HaC filter is an [N~II] $\lambda$6584 filter. NGC~6778 was also imaged as part of our FORS2 observations for 60 s with the OIII filter (500.1/5.7 nm). In the NGC~6326 images (0.146\arcsec/pixel) the average seeing ranged between 0.76\arcsec\ and 0.79\arcsec\ (Ha and HaC), and 0.92\arcsec\ and 0.97\arcsec\ (OIIIC and OIII), while 0.86\arcsec\ was measured for the FORS2 NGC~6778 [O~III] image (0.252\arcsec/pixel).
   Marked in Fig. \ref{fig:montage} are the CSPN positions and the tips of collimated outflows. We refer the reader to Miranda et al. (2010) for sub-arcsecond imaging of NGC~6778 that best reveals its extreme filamentary nature.

   \section{Nebula morphologies}
   \label{sec:discussion}
   \subsection{General properties}
   \label{sec:morph}
   The basic underlying shape of NGC~6326 can be gleaned from the GMOS OIII and OIIIC images showing an elliptical nebula, while NGC~6778 is unambiguously a bipolar nebula (Maestro, Guerrero \& Miranda 2004; Miranda et al. 2010) with a prominent torus and minor axis at a position angle (PA) of 107$^\circ$. The torus of NGC~6778 is notable for its high expansion velocity of 2$V_\mathrm{exp}$=60--70 km/s (Maestro et al. 2004), while Meatheringham, Wood \& Faulkner (1988) found a slower 2$V_\mathrm{exp}=30\pm3$ km/s for NGC~6326. It is interesting to note that other post-CE nebulae show similarly high expansion speeds as NGC~6778 including ETHOS~1 (2$V_\mathrm{exp}$=110 km/s, Miszalski et al. 2011b), A~41 (2$V_\mathrm{exp}$=80 km/s, Jones et al. 2010) and the Necklace (2$V_\mathrm{exp}$=56 km/s, Corradi et al. 2011).
   
   Both objects exhibit collimated outflows or jets, possibly launched via a short-lived accretion disk or CE dynamo, as would be expected for post-CE nebulae (Soker \& Livio 1994; Nordhaus \& Blackman 2006). They add to the mounting observational evidence of jets associated with post-CE PNe (Mitchell et al. 2007; Miszalski et al. 2009b; Corradi et al. 2011; Miszalski et al. 2011b). In NGC~6326 we have noted for the first time a thin pair of jets at PA=145$^\circ$ that complements what seems to be jets beyond the `whiskers' at PA=50--80$^\circ$. High resolution spectroscopy of NGC~6326 is planned in the near future to check this interpretation and quantify the kinematics of these jets with a full reconstruction of the nebula. The fact that NGC~6778 is eclipsing (Fig. \ref{fig:lc}) is consistent with the nebula inclination of $i\sim85^\circ$ derived by Maestro et al. (2004) and aligns the pair of jets at PA=15$^\circ$ with the polar axis of the nebula.
   
   \subsection{Low-ionisation filaments} 
   Most remarkable in both nebulae is their rich attribution of filamentary low-ionisation structures (LIS) whose origin has long been uncertain (e.g. Gon{\c c}alves et al. 2001; Corradi 2006). Miszalski et al. (2009b) suggested close binaries play a strong role in the formation of LIS after finding an elevated occurrence of LIS in post-CE nebulae. Adding NGC~6326 and NGC~6778 further strengthens this growing connection which may be explained by the photoionisation of neutral material deposited in the orbital plane during the CE phase (e.g. Raga, Steffen \& Gonz\'alez 2005). Similarly rich levels of filaments are found in the post-CE PNe NGC~6337 (Corradi et al. 2000), K~1-2 (Corradi et al. 1999) and Longmore 16 (Frew et al. in prep). Sparser rings may also be found in Sab~41 (Miszalski et al. 2009b) and the Necklace (Corradi et al. 2011). 
   
\begin{figure}
      \begin{center}
         \includegraphics[scale=0.4]{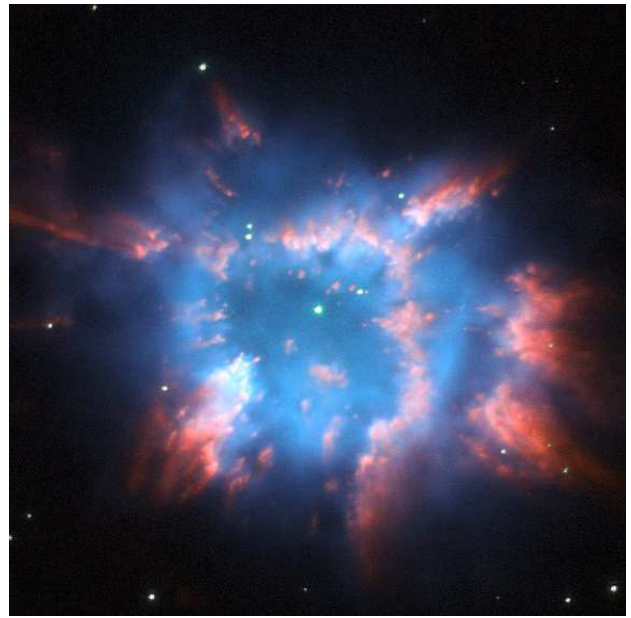}
      \end{center}
      \caption{\emph{HST} colour-composite image of NGC~6326 made from archival images taken in F658N (red), F555W (green) and F502N (blue) filters. Image credit: ESA/Hubble and NASA. }
      \label{fig:N6326hst}
   \end{figure}

   Figure \ref{fig:N6326hst} best reveals the cometary, fragmented tails of LIS in NGC~6326, the most linear of which all point back to the CSPN position. The public domain \emph{HST} image\footnote{http://spacetelescope.org/images/potw1010a} is made from archival WFPC2 exposures of 1400 s in F658N, 360 s in F502N and 260 s in F555W. Such a stark disconnect between the relatively smooth elliptical main nebula and its `spiky' filaments is not commonly observed in PNe. Very similar filaments appear in the PNe A~30 and A~78 (Borkowski et al. 1993,1995). In our case the filaments are neither H-deficient, nor expanding at nova-like speeds, however there is a strong morphological similarity. Lau, De Marco \& Liu (2011) suggested that A~30, and the related A~58, could potentially be the result of CE evolution rather than so-called `born-again' evolution. Our work provides additional circumstantial evidence in support of this argument. There is also a superficial level of resemblance with classical nova shells, known for being fragmented into hundreds of knots (e.g. Shara et al. 1997; Bode, O'Brien \& Simpson 2004), but in our case the low expansion velocities preclude a thermonuclear origin. Nova shells also retain a more well defined spherical or ellipsoidal geometry compared to the more chaotic cometary filaments with no overall governing geometry seen in PNe. 
   
   \section{Conclusions}
   \label{sec:conclusion}
   We presented central star spectroscopy and lightcurves of the bright Southern PNe NGC~6326 and NGC~6778 that proves their close binary nature with main-sequence companions in orbital periods of 0.372 and 0.1534 days, respectively. The combination of low-ionisation filaments and multiple collimated outflows (jets) in their nebulae further supports the suspected association between these features and post-CE binaries (Miszalski et al. 2009b). A wealth of low-ionisation filaments in NGC~6326 in particular is comparable to filaments in A~30 and A~78, which adds circumstantial evidence to the case for at least A~30 having post-CE CSPN (Lau et al. 2011). The eclipsing CSPN of NGC~6778 is consistent with the nebula inclination determined by Maestro et al. (2004) and has one of the shortest orbital periods known for PNe. 

   NGC~6326 is to our knowledge the first post-CE CSPN discovered spectroscopically before an orbital period was later obtained after photometric monitoring, while our observations of NGC~6778 are essentially contemporaneous. Given absorption lines are susceptible to intrinisic non-orbital variability (De Marco 2009), the C~III and N~III emission lines are a powerful alternative window into detecting binarity in PNe. Their large RV shifts of 100 km/s, relative to nebula lines, make them accessible to 8-m class telescopes with intermediate resolution spectrographs. On smaller telescopes their strength means lower resolution time-domain spectroscopy could potentially reveal the tips of these lines appearing and disappearing as in Fig. \ref{fig:fors}. Such surveys are ideally suited to pre-select binaries for more time consuming photometric monitoring and will be able to routinely access CSPN of brighter nebulae that are underrepresented in known post-CE nebulae. 

   Those CSPN classified as \emph{wels} are an excellent sample to start with, since they simply have the CIII/NIII lines to probe for RV shifts, and we expect many to turn out to be close binaries (Miszalski et al. 2011b). The actual number of binaries expected to be found is poorly constrained at present as no large problem-specific surveys have been conducted. Of course, by `many' we do not mean all \emph{wels} will be binaries simply because of the large inhomogeneity of the `class' (Marcolino \& de Ara{\'u}jo 2003). There are currently no surveys which specifically target \emph{wels} spectroscopically and few have been monitored photometrically to gauge the actual binary fraction of the heterogeneous \emph{wels} sample. Hajduk et al. (2010) obtained less than 15 $I$-band observations of 5 \emph{wels} CSPN to gauge their variability and found no variables. This is understandable given the expected orbital periods may reach up to 1--3 days (Miszalski et al. 2011a) and cannot be probed by undersampled observations of $\sim$1--2 hours (no matter how carefully spaced since the period is \emph{a priori} unknown). It is also difficult to judge variability based on the photometric scatter alone since at least a third of close binary CSPN may be expected to have amplitudes $<$ 0.1 mag (Miszalski et al. 2009a). A further 5 \emph{wels} listed by Hajduk et al. (2010) were also studied by Miszalski et al. (2009a) who did not find any periodic variability. These findings are not significant enough to indicate one way or the other whether \emph{wels} are associated with close binaries and much further work is needed in this area. 
   Classification schemes of central stars should avoid applying the obsolete \emph{wels} label once an object has been proven to be a close irradiated binary. Central stars that fall into this overlap in classification schemes include, amongst others, NGC~6326, ETHOS~1 (Miszalski et al. 2011b), the Necklace (Corradi et al. 2011) and A 46 (Pollacco \& Bell 1994).
  
\begin{acknowledgements}
    BM acknowledges the conscientious and helpful support of Simon O'Toole and Steve Margheim during the course of Gemini program GS-2009A-Q-35. MS-G acknowledges support by the Spanish MICINN within the program CONSOLIDER INGENIO 2010, under grant “ASTROMOL” (CSD2009-00038). We thank an anonymous referee for comments that helped improve this work. This paper includes observations made at the South African Astronomical Observatory (SAAO), the Roque del Los Muchachos observatory, Paranal Observatory, as well as observations made at the Gemini Observatory which is operated by the Association of Universities for Research in Astronomy, Inc., under a cooperative agreement with the NSF on behalf of the Gemini partnership: the National Science Foundation (United States), the Science and Technology Facilities Council (United Kingdom), the National Research Council (Canada), CONICYT (Chile), the Australian Research Council (Australia), Minist\'erio da Ci\^encia e Tecnologia (Brazil) and Ministerio de Ciencia, Tecnolog\'ia e Innovaci\'on Productiva (Argentina). 
\end{acknowledgements}

\end{document}